# The Peaks of Eternal Light: a Near-term Property Issue on the Moon

Martin Elvis [a,1], Tony Milligan [b], Alanna Krolikowski[c]

[a] Harvard-Smithsonian Center for Astrophysics, 60 Garden St., Cambridge MA 02138, USA; melvis@cfa.harvard.edu;
[b] Department of Theology and Religious Studies, King's College London, Virginia Wolf Building, 22 Kingsway, London WC2B 6NR.
[c] Georg-August Universität Göttingen, Heinrich-Düker-Weg 14, 37073, Göttingen, Germany; alanna.krolikowski@gmail.com

**ABSTRACT**

*The Outer Space Treaty makes it clear that the Moon is the 'province of all mankind', with the latter ordinarily understood to exclude state or private appropriation of any portion of its surface. However, there are indeterminacies in the Treaty and in space law generally over the issue of appropriation. These indeterminacies might permit a close approximation to a property claim or some manner of 'quasi-property'. The recently revealed highly inhomogeneous distribution of lunar resources changes the context of these issues. We illustrate this altered situation by considering the Peaks of Eternal Light. They occupy about one square kilometer of the lunar surface. We consider a thought experiment in which a Solar telescope is placed on one of the Peaks of Eternal Light at the lunar South pole for scientific research. Its operation would require non-disturbance, and hence that the Peak remain unvisited by others, effectively establishing a claim of protective exclusion and de facto appropriation. Such a telescope would be relatively easy to emplace with today's technology and so poses a near-term property issue on the Moon. While effective appropriation of a Peak might proceed without raising some of the familiar problems associated with commercial development (especially lunar mining), the possibility of such appropriation nonetheless raises some significant issues concerning justice and the safeguarding of scientific practice on the lunar surface. We consider this issue from scientific, technical, ethical and policy viewpoints.*

## 1. INTRODUCTION

Existing international agreements, the Outer Space Treaty (OST) in particular, seem to exclude the appropriation of the Moon and other celestial bodies by any terrestrial state and, by implication, private appropriation also. Under the OST, space is the 'province of all mankind'. Yet there are clear pressures towards the recognition of property rights in space [1]. The most obvious example here is the *Space Resource Exploration and Utilization Act* passed by the US Congress in 2015. [2] The possible introduction of a wider regime of property rights has also been much discussed [2,3,4]. Disagreement over the likely shape of any future rights regime is strong and the arguments do not look like they will be resolved until one or more practical examples of appropriation for use are at hand. Such practical examples have seemed far off. Here, however, we suggest an example that could become urgent in within years rather than decades.

---

[1] Corresponding author.
[2] https://www.congress.gov/114/bills/hr1508/BILLS-114hr1508rh.pdf

Because the issue of legality is to the fore, property rights on the Moon are often discussed against the backdrop of a convenient fiction, an assumption that the Moon is more or less uniform and that, for the most part, occupying one region of the Moon would not deprive others of lunar resources. However, in his recent review of the latter, Crawford (2015) [5] brings together the latest lunar maps from a wide range of sensors and these show a highly structured surface at kilometer scales. They show the extent to which lunar resources are not uniformly distributed. Appropriation by one agent or group of agents could deprive others of important opportunities, and could thereby generate a situation of injustice. A favorite example of a relatively rare resource is water, which may persist in permanently dark craters [6], which in turn occupy thousands of square kilometers in the polar regions (Figure 1 [7]) [8,9]. Even so, they still occupy only a small fraction (0.1%) of the total lunar surface.

Scarcer still are the regions of permanent solar illumination, the "peaks of eternal light", described in section 2. If such peaks can be found near to permanently dark craters then we have hit the lunar resource jackpot: a region with great resources next to a region with abundant power. Just as the proximity of iron and coal mines in England was an important spur to the Industrial Revolution [10], so this combination of power plus resource may spark or at least facilitate industrial development on the Moon. As we will see this combination is also extremely rare on the Moon and it is the power supply that is scarcest of the pair.

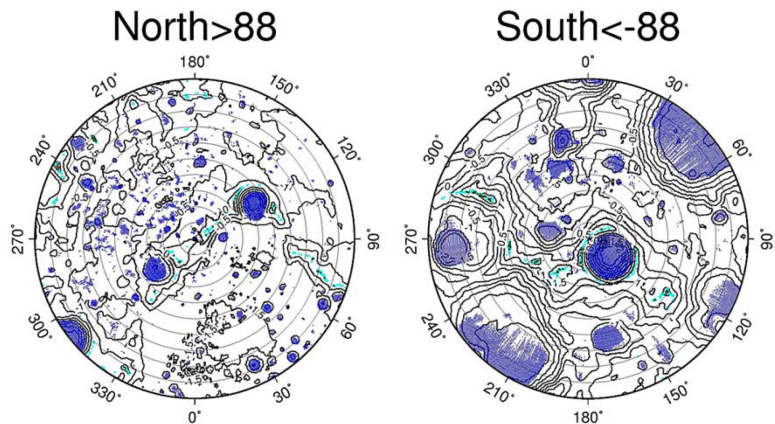

**Figure 1:** Lunar Polar regions (88-90 degrees) showing zero illumination regions (purple) and high illumination regions (light blue) [Source: [7] Noda et al. 2008].

As always when resources are concentrated, this clustered distribution of high illumination regions will eventually lead to disputes over rights to those resources. And the scarcer (and more valuable) the resource, the more eager we may expect agents to be in staking their claim once the resource is believed to have been reliably identified. The first case of trusted identification might trigger a "scramble for the Moon" comparable in some respects to the "scramble for Africa" which began with the identification of mineral resources in the Congo in the 1880s [11]. And here, the appeal to the African precedent may be a better fit than any appeal to the 19th century American frontier, given the high level of resources, and state-led activity which the former required by contrast with the appropriation of resources by small groups of individuals in the American West. Given the growing realization that the Moon does have scarce concentrations of resources, the issue of appropriation, which has appeared distant, may soon emerge as a more practical problem in urgent need of solution.

In the case of the peaks of eternal light the resource in question is so scarce (roughly 1/100 of a billionth of the lunar area) that even a single country or company could, on its own, occupy them all, effectively denying that resource to others. In what follows, we show how scarce a resource they are, how they could be (in some sense) legally appropriated using ambiguities in the Outer Space Treaty to uphold such an appropriation. We then consider how to deal with this near-term rights issue on the Moon.

## 2. PEAKS OF ETERNAL LIGHT

The "Peaks of Eternal Light" are highland regions near the lunar poles that receive sunlight virtually all of the time. I.e. they are (almost) never shadowed by other parts of the Moon. The existence of such peaks was first hypothesized by Beer and Mädler (1837, p.16) [12]. Over 40 years later, the popular science writer, Camille Flammarion gave them their poetic name: "montagnes de l'éternelle lumière" (translated as "peaks of eternal light") [13,14]. The small tilt of the Moon's spin axis to the ecliptic (1.54° versus 23.5° for the Earth) makes these peaks possible. In the 19th century it was not clear if any such peaks actually existed.

Searches for regions of nearly permanent sunlight have, however, become a priority for lunar explorers as they are excellent places to situate scientific experiments or human habitats. By providing a virtually permanent source of power from sunlight, equipment emplaced there can be simplified. The main advantages are that fewer high mass batteries need to be transported to the Moon to provide power during darkness, as periods of darkness are short, and that the rigors of the low temperature lunar night are mostly avoided, thereby simplifying instrument and habitat design.

The existence of peaks of eternal light is no longer conjectural. Over the past decade several lunar-orbiting spacecraft (SMART-1, Clementine, SELENE, Chandrayaan, Lunar Reconnaissance Orbiter) have mapped the lunar poles in increasing detail [5]. Some candidate "eternal light" regions near both the North and South lunar poles were found from these maps. An illumination map of the North polar region is shown in Figure 2 [15]. The red and white areas have 90-99% and 100% illumination respectively. Note how thin these regions are. They are primarily ridges and crater rims.

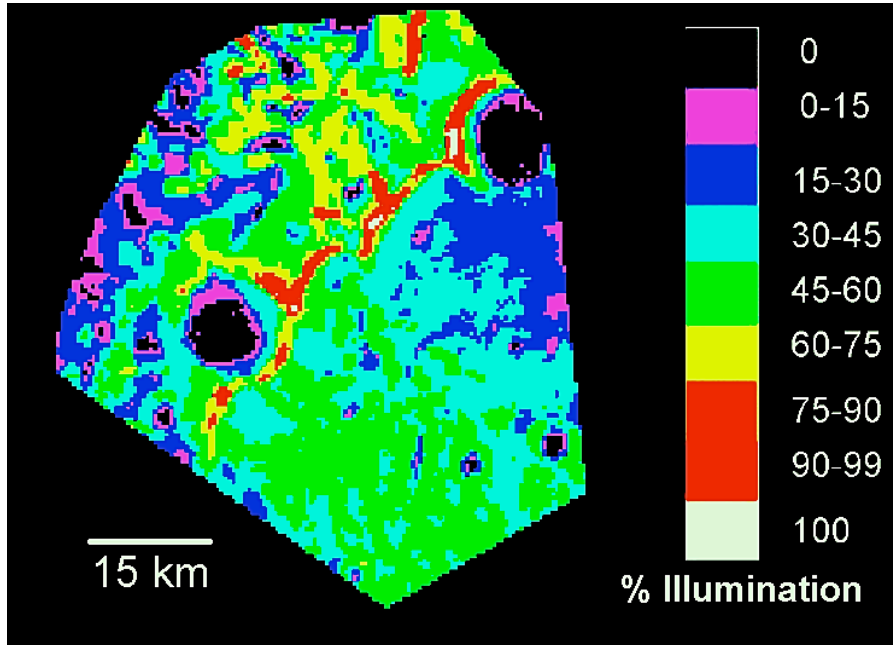

**Figure 2**: Illumination map of the Moon's north pole, within about 30 km (1-1.5 degrees). The narrow strips of orange, red and white are illuminated at least 75% of the time. Source: Bussey et al. (2005) [15].

The South lunar pole is a rather different matter. While the North Pole happens to be a highland region, the South Pole lies in the large (2500 km diameter) South Pole Aitken impact basin, several kilometers below the mean lunar surface. Bussey et al. (2005) [15] say that at the South lunar pole there is no permanently illuminated region, at least on the 500m scale that they can measure. As a result there are no true "eternal light" regions at the south lunar pole. There are still, however, regions of high illumination [16]. Like the true "eternal light" regions in the north, these high illumination regions in the south are small, as seen in the topographic simulation shown in Figure 3. The total area of high illumination, where the surface sees sunlight over 70% of the time, are each only a few hundred square meters. For comparison a FIFA regulation size soccer pitch is 100-110m long by 64-7m wide. Each, typically elongated, area is then equivalent to a few meter wide strip along the length of a soccer pitch.

However, the situation is not quite as poor at this number suggests. Gläser et al. (2014) [16] also looked at the effect of altitude above the surface on the illumination. Even raising the surface by 2m makes a difference; for a 10m raise it is possible to find a 200 m × 500 m region with >90% illumination. Even then, the peaks of eternal light (hereafter PELs) are only slivers of the Moon's surface. This extreme scarcity combines with an interesting tension with the OST.

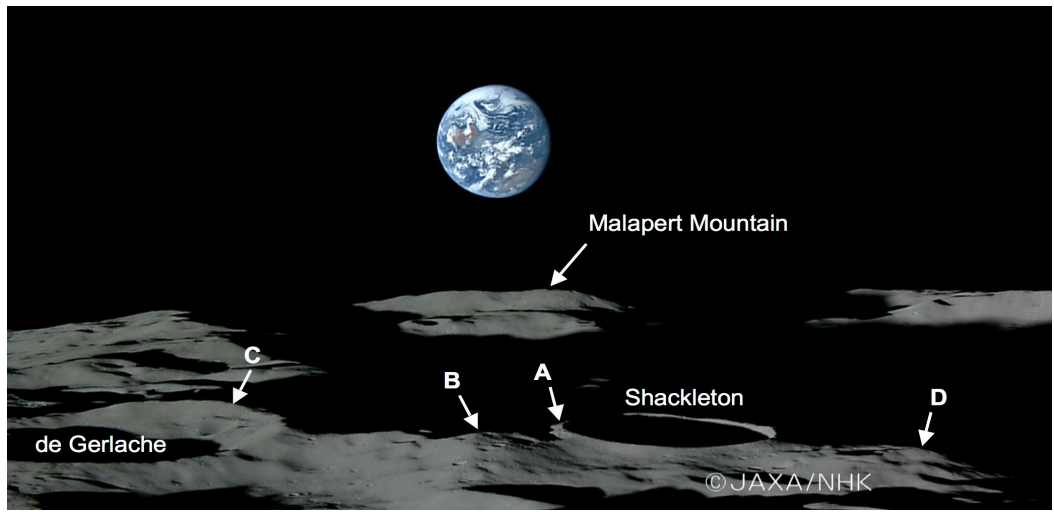

**Figure 3:** Lunar South Pole, four peaks are identified which are illuminated more than 80% of the time. (source: Bussey et al. 2010 [17]).

## 3. A TENSION WITHIN THE OUTER SPACE TREATY

The Outer Space Treaty[3] (OST) Article II states (in full) that: "*Outer space, including the moon and other celestial bodies, is not subject to national appropriation by claim of sovereignty, by means of use or occupation, or by any other means*." This seems clear, and it is not our intention to play upon the familiar skewing of the dominant sense of the Treaty by appealing to the fact that the envisaged appropriating agents, at the time of its drafting, were nations rather than private commercial enterprises. Both are, explicitly or by implication, excluded in favor of a shared claim made on behalf of humanity at large.

However, OST Article XII states (in full) that:

*"All stations, installations, equipment and space vehicles on the moon and other celestial bodies shall be open to representatives of other States Parties to the Treaty on a basis of reciprocity. Such representatives shall give reasonable advance notice of a projected visit, in order that appropriate consultations may be held and that maximum precautions may be taken to assure safety and to avoid interference with normal operations in the facility to be visited."*

There are (at least) three background principles brought into play here, principles of *openness*, *reciprocity* and *precaution*. How best to formulate and prioritize these principles is unclear. It does seem that it will be difficult to do so in a way that remains faithful to the text while avoiding all possible cases of conflict between the principles themselves. Indeed, the appeals to openness and reciprocity on the one hand and precaution on the other, do not sit well together. Openness and reciprocity suggest that there shall be a system of access to all areas and facilities. However, if a country, or organization, were to set up a delicate scientific experiment it might be impossible for any visitors to approach it without interfering with its normal operations. The precautionary principle, at least on a strong reading, would then clash with the other principles. Experimental integrity would clash with openness and with reciprocity as well (assuming that the agents conducting the experiment would still retain their entitlements with regard to the sites of other agents operating elsewhere on the lunar surface).

---

[3] Or, formally, the "Treaty on Principles Governing the Activities of States in the Exploration and Use of Outer Space, including the Moon and Other Celestial Bodies".
(http://www.unoosa.org/oosa/SpaceLaw/outerspt.html, accessed 2015 May 18.)

If the precautionary principle were regarded as trumps then, in all but name, the experiment site would have become effectively "owned" by the country setting up the experiment or, if not owned, at least legally appropriated. A de facto "quasi-property" status might well be accepted, especially if there were no push to formalize matters. At the PELs though, one scientific appropriation could pre-empt another. E.g. the area would not be available to provide solar power for an exploration of the permanently dark craters nearby. Nor would it be available for any industrial use of their water deposits. (And here, there are also likely to be disagreements about the respective importance of lunar science and any prospective industrial exploitation.)

True, such appropriation would not erase all occupancy-related obligations and requirements, but claims of property are themselves not absolute in this regard. They too are subject to all manner of legal constraints. Property comes with liabilities and vulnerabilities to legal seizure. The suggestion, then, is not that existing international treaties might be circumvented in order to set up an effective claim (in the manner of Dinkin, 2004 [18]) but rather that when a precautionary principle was introduced into space law, it was not formulated with an eye towards the grey boundary which separates out property from non-property, a boundary which is far less sharp than we might imagine.

Two problems then seem to emerge. Firstly, is this a plausible or merely convenient reading of the OST? Secondly, is there any realistic experiment that could bring the precautionary principle into play in an authoritative manner? (If the experiment was deemed to be pseudo-science any appeal to the precautionary principle might then carry little weight.) Additionally, we may wonder whether or not the setting up of such an experiment could be feasible in the short-term, making its consideration urgent.

The answer to the first question may depend upon broader issues in jurisprudence. If, for instance, we hold that the law is always determinate, that there is always 'one right answer' even on unanticipated matters [19,20] we might then simply deny that the alleged ambiguity is present and resolve the matter in some way which clearly rules out any effective appropriation, even without the formal title of lunar property. What makes this a tough argument is the extent to which it clashes with the regular practice of law, a good deal of which seems to involve the search for loopholes and appeals to judicial discretion. Arguments for uniform legal determinacy have always looked suspiciously like a presupposition about the nature, and specifically the reach, of law, one that requires us to re-describe actual legal practice which persistently, and stubbornly refuses to conform.

If, by contrast, we accept that the apparent ambiguities of law do not always disappear under further analysis [21] then it may be difficult to deny that the OST really is indeterminate over issues of how much protection an experimental site is due, for how long the protection should remain in force and what form it should take. A country, or group of agents held responsible to a country as their launch state, might effectively take possession of an "eternal light" area and then argue what could well be a plausible case for unhindered occupancy. This would not yield property in the sense beloved of 18[th] century legal theory (with an overstated security of possession), but it would yield everything that matters in terms of appropriation and effective control.

## 4. A THOUGHT EXPERIMENT INVOLVING APPROPRIATION FOR SCIENTIFIC USE

The answer to the second question is almost certainly 'yes'. There is at least one non-trivial use of a "Peak of Eternal Light", to which the latter are peculiarly suited, and which requires the equipment involved to be free of intrusive activity.

Let us imagine that a simple radio telescope is set up on the south pole high illumination regions. Radio astronomy is an obvious candidate here because it is the only astronomy discipline that would benefit from being located on the Moon. The normal location proposed for lunar radio telescopes is on the far side, where interference from our terrestrial radio emission would be minimal (see chapters by Erickson, Weiler, Dwarakanath, Woan, Duric and Kassim & Yusef-Zadeh in [22]). There is also perhaps the last unexplored region of the electromagnetic spectrum available in space: long wavelength radio. The Earth's atmosphere is effectively opaque to radio waves that have wavelengths longer than about 10 m [22]. This is equivalent to a frequency of 30 MHz.

The polar high illumination location would be needed to study the Sun *without interruption*. The absence of the latter matters because the technique of Fourier analysis of variability in a signal is greatly simplified if continuous data is available. If the radio antenna were held up a few meters above ground level, then the south pole location would work well. This would be a Solar radio astronomy telescope to provide *continuous* monitoring of the Sun at wavelengths longer than 10m. Chapters by Robinson & Cairns; Bastian; Rickett & Coles in [22] discuss the value of solar observations in this regime.

The telescope would consist of a dipole antenna, essentially a long uncovered wire. To have any useful angular discrimination it would need to be 100 m long or more. This requirement is due to diffraction: a 100m long antenna detecting 10m long radio waves could tell their direction to within about 6 degrees. (I.e. about 6 times the Sun's diameter as seen from the Moon.) Hence it would naturally run the length of one of the long, thin, Peaks of Eternal Light.

Unspooling a long wire using a small rover would be a relatively simple, but usefully challenging operation. Putting the antenna wire on posts several meters high would involve another level of challenge. This radio telescope could then be correctly presented as both valuable scientifically in itself, and as a precursor for more ambitious far side telescopes. Of course the north pole would still at least appear to be a better location, as we saw above that is has true peaks of eternal light, and areas of illumination are larger than those at the south pole. However, given the greater immediate interest in exploring the south pole, it may be that this location could be justified by precursor missions that map out the territory at the fine level of detail (centimeters) to make the installation possible.

Installation of such a simple solar radio telescope is entirely feasible within a few years rather than decades (although, exactly how few is a matter for debate). Several countries are already discussing lunar landers.[4] China has recently (in 2013) put the *Chang'e 3* lander and *Yutu* rover onto the Moon, demonstrating most of the required technology. *Chang'e 4* may land near the Aitken basin or on the far side. Japan's JAXA space agency announced lunar lander plans in 2015.[5] Private companies are also claiming a near similar level of technological readiness, some inspired by the Google Lunar X-prize.[6] Others, such as Moon Express and Shackleton Energy, are motivated by more direct commercial considerations.[7] All are expecting lunar landings in the next few years, including landings near the lunar South pole, and quite likely on the Peaks of Eternal Light.

A radio telescope of this type would be extremely sensitive to electrical interference from any electrical equipment in its vicinity. The posts that support the antenna would also be vulnerable to mechanical disturbance, as they will likely not be strongly anchored to the lunar surface regolith. Hence any other equipment that approaches this experiment would be interfering with its "normal operations". And so, while Article XII of the OST states that all *"… equipment… shall be open to… other states"* the nature of the simple solar radio telescope we described above obviates the very possibility of any site visits without causing an interruption of the data or destruction of the experiment.

Effectively a single wire could co-opt one of the most valuable pieces of territory on the Moon into something approaching real-estate, giving the occupant a good deal of leverage even if their primary objective was not scientific enquiry. Our point here has the additional virtue of being independent of any claims about special acts of 'original appropriation' that might, because of the nature of the acts themselves, carry the potential to transform non-property into property, and independent also of the associated (and disputable) philosophical frameworks about how anything first becomes property. We can remain official neutral about such matters and still generate a significant problem of effective co-opting. Strategic occupancy might then be negotiated into concessions with regard to other lunar activities and might impact upon how space law (including the OST) was subsequently read.

---

[4] Space Daily, 22 May 2015. 'China Plans First Ever Landing On The Lunar Far Side'. http://www.spacedaily.com/reports/China_Plans_First_Ever_Landing_on_the_Dark_Side_of_the_Moon_999.html
[5] http://www.cnn.com/2015/04/23/tech/japan-moon-lander-planned/ Accessed 15/07/2015.
[6] http://lunar.xprize.org Accessed 15/07/2015.
[7] http://www.moonexpress.com Accessed 15/07/2015; http://www.shackletonenergy.com/ Accessed 15/07/2015.

Clearly this state of affairs offers a situation of possible strategic advantage that would be disputed by others. Article I forbids sovereignty claims based on "*use or occupation*". But the state or agency that sets up the experiment would not actually need to claim sovereignty (and might be ill-advised to do so); it would simply require that others abide by their legal and precautionary obligation not to interfere with normal operations. Whether this is *effectively* or *to all intents and practical purposes* claiming sovereignty would be an open question, something that might be argued about.

OST Article IX does say that:

*A State Party to the Treaty which has reason to believe that an activity or experiment planned by another State Party in outer space, including the moon and other celestial bodies, would cause potentially harmful interference with activities in the peaceful exploration and use of outer space, including the moon and other celestial bodies, may request consultation concerning the activity or experiment.*

As soon as the experiment is put in place others would, no doubt, raise exactly this objection. As of now, however, there is no mechanism for resolving such a dispute and no remedy for anyone claiming that their rights have been violated or illegitimately compromised by such an action.

## 5. OCCUPANCY AND JUSTICE

This example is a useful *gedankenexperiment* as it brings the rather abstract notion of property rights in space "down to Earth" and into an immediately practical setting. Nor does it describe any singular circumstance of the sort that is likely to remain strictly hypothetical. It can also be used to support a more general claim about the likely emergence of disputes that do not admit of any easy resolution under existing space law. It is likely that as we learn more about space resources we shall find many other cases where their vast abundance may lead us to overestimate the small number of *accessible* locations where they may actually be found (e.g. near-Earth asteroids, [23]). As a result, these resources will almost certainly become extremely valuable and subject to disputes about occupancy and use.

A concern about over-regulation as a possible brake upon commercial engagement with outer space might lead us to accept this situation without further, anticipatory, action. We might even, in extremis, favor some form of celestial anarchy with disputes about lunar territory settled more or less ad hoc [24]. The disadvantages of such a situation include those which shaped the initial framing of the OST, most especially the prospect of worsening the already fragile relations between those terrestrial states which are the main players. The current revival of new forms of political tension between the main players in space is a reminder that the terrestrial impact of rivalries in space should be an ongoing concern.

This alone could lead us to favor the idea of building a framework for regulation, although it could be difficult to do so. Future agreements might take a form closer to bilateral agreements than a new treaty that could be ratified by all the main parties concerned. The OST was, after all, the product of a very specific configuration of political circumstances at a time when no-one had actually set foot on the Moon. The avoidance of advantage to others was a key consideration. It is at least conceivable that the days of such great treaties are over. As an alternative, it might be best to pursue a consensus or even a rough agreement about how the existing treaties ought to be read: i.e. what falls within the bounds of a 'best fit' and 'best light' reading, and how certain kinds of ambiguity about occupancy and resource use are best resolved, in the least tension-producing manner.

Part of the discussion about how that might be done will concern *realpolitik*, but the discussion could, and perhaps also *should,* bring a number of ethical considerations into play. It is, at least, difficult to understand what might be involved in reading existing space law in the 'best light' if not an appeal to current moral standards together with ethical and social considerations of a broader sort. This move need not collapse the important distinction between ethics and law, but it does affirm the simpler point that the distinction is not a rigid dichotomy. Law is, in many respects, a normative practice. That is to say, it concerns what *ought* to be done. And where the law turns out to be ambiguous, a resolution of the ambiguity in line with prevailing social and political values and with prevailing ethical commitments has a clear role to play. This is part of what it is to consider law in the context of a democratic society, as an institution that is answerable to society at large, rather than simply regarding law as a system of commands.

There are four obvious candidate ethical considerations that may help us to make sense of what an *appropriate* legal and regulative framework for PEL occupation could look like. All are well represented in the existing literature on the ethics of human activity in space. The ordering below moves from ethical considerations which may be harder to embody in law, towards those which are already, to some extent, included.

(1) We may have a duty to extend our human presence out into the universe and *any* regulative framework for the Moon needs to fit in with this goal [4, 25].

(2) We may have a duty to extend the boundaries of scientific knowledge [26]. Effective occupation of an "eternal light" area could block this.

(3) We may have duties concerning planetary protection [28].

(4) We may have duties concerning the just distribution of opportunities and resources [27].

Even if we are sympathetic to (1), it is likely to be an ineffective driver for the shaping of our legislative framework, or for a 'best light' reinterpretation of the latter. Firstly, such a duty may be moot because it is not obvious that we *can* significantly extend our human presence in the way envisaged. While we may not exactly be stuck here on the Earth, our movements may be restricted to nearby regions of space. The distances we would have to travel to reach anywhere else may simply be too great. Secondly, even if the relevant expansion is possible, it may be such a long-range consideration that we have no real way of knowing whether a particular series of actions is ultimately likely to help or to hinder. We just do not know enough to guide decisions about rights of access or legal and policy frameworks more generally. Appealing to such a duty is then, at best, a weak tool in the present context.

Consideration (2) is rather more likely to be action guiding. There is the possibility that limited and valuable resources might be used or monopolized for trivial purposes or, at least, for less than maximal purposes. Worthwhile and important science activities might get squeezed out by something rather less worthwhile, especially given the connection between lunar activity and political prestige. It is not obvious, for example, that the emerging Chinese lunar program is driven more by scientific curiosity than by political profile on the terrestrial world stage. (In this respect, it follows a long line of Western precedents.) Geopolitically, the sheer fact that a nation can say that it is conducting experiments on the Moon may be more important than the kind of experiments that it is actually conducting. Given this, there may be strong ethical grounds for the protection of scientific opportunity on the "peaks of eternal light".

How, though, are the maximal or higher purposes ever to be agreed upon? Even within science this is likely to be contentious. Yet there may be the prospect of securing a consensus on at least some matters. One terrestrial analogue here would be whaling under the guise of scientific enquiry in order to provide whale meat for the domestic Japanese market. There is a fairly broad consensus both that the science is a pretext and that it has little value. In March 2014 the International Court of Justice ruled that such whaling was not for scientific purposes and ordered its cessation.[8] The point here is not that every case will be quite so simple, but rather that a workable consensus on what constitutes valuable science *can* sometimes be achieved even though it may take some time to achieve it.

What might ease the way for regulation, or at least for an agreed 'best light' reading of the OST and of the associated treaties, is that considerations (3) and (4) are already embedded in the existing legislation. Albeit, the "planetary protection" in (3) is ordinarily taken to concern only protection for reasons of scientific and cultural importance rather than protection of the Moon and other bodies per se, as having some manner of inherent value in their own right [28]. On the plus side, unlike lunar mining, PEL occupation (unless it is in some way linked to more intrusive processes) is unlikely to raise major environmental damage concerns. There is no prospect, as there would be with He3 mining, of vast areas of lunar regolith being scraped up, year by year, for processing. Indeed, from the standpoint of lunar protection (however understood) PEL occupation for astronomical observation might be a very good thing precisely because it would effectively create protected zones where intrusive activities would be in violation of the precautionary principle that the OST affirms.

There is, however, no getting around the problems of distributive justice which are likely to result from any effective appropriate of such a limited resource. This applies to the peaks of eternal light, just as it would in the case of any plausible candidate for asteroid mining or the occupation of prime orbital niches around the Earth [27]. When a resource to which we have a shared entitlement is monopolized, something has gone wrong. Analogously, the first and most powerful guests at a party do not get to claim all of the available food and drink, even if they have fast cars which ensure that they arrive before anyone else. We might, however, wonder if such analogies are the best ones available. The Californian gold rush of 1849 did herald the introduction of first-user claims upon water in the American West [29]. Interestingly, to establish a legitimate prior-appropriation claim, the appropriation had to be for 'beneficial use' (which is just as open to argument and interpretation as discussions of what constitutes valuable science). However, overall, first use is not viable. California operates a dual system of riparian rights (permitting multiple use) and the prior-appropriation/first-use doctrine together with residual tribal rights.[9] Recent drought conditions in California have also led to speculation about how long first-use claims can reasonably be sustained, as part of this mix, to the overall detriment of the population at large.

---

[8] For the March 2012 International Court of Justice ruling that whaling was not being carried out for scientific purposes see http://www.icj-cij.org/docket/index.php?p1=3&p2=1&case=148&code=aj&p3=4

In addition, an advocate for first user rights on the Moon should pause to consider the fact that PELs are a tiny part of the Moon's resources. It is not obvious that the first user of PELs will be the first user of other lunar resources. And the resources of the Solar System, beginning with the asteroids, are far larger still. The first user advocate would be well advised to exercise caution as they are operating behind a "veil of ignorance" [30], not knowing whether this user will be themselves or someone else, and not knowing the overall distribution of resource access [31]. A poorly chosen precedent could see them occupying peaks of eternal light but having access to little or nothing else.

On this matter, there is a close alignment of the OST with considerations of just distribution (4). The treaty is quite clear that parties who have a claim upon space do so equally and this entails an *ongoing* commitment to distributive justice. Accordingly, with regard to the PELs, the treaty poses familiar problems in relation to the states or segments of humanity who will be excluded from equal use if the peaks have already been seized or, more simply, because the disadvantaged groups happen to lack access to space or to the Moon.

This preliminary review of the background ethical issues suggests that the primary ethical concerns which should be in play if we are to reshape the legislative framework appropriately in line with the issues raised by the PELs are (2) the protection of science and (4) considerations of fair and reasonable distribution of limited lunar opportunities and resources. Given the likely non-intrusive nature of PEL occupation, consideration (3) is of relevance in the sense that it is relevant to all human activity on the Moon but unlike (2) and (4) it has no special relevance peculiar to the issue of PELs. Consideration (1) is, as stated before, unlikely to be action-guiding on this matter.

Although this suggests that distributive justice over PEL occupation and access is likely to be one of the two main ethical consideration, we need not appeal to any broader context of distributive justice and how human activity on the Moon might fit into it. Issues of fairness in relation to lunar access might fit well into a broader, redistributive social agenda but they need not be thought of as part of the latter in order for their significance to be acknowledged.

---

[9] http://www.swrcb.ca.gov/waterrights/board_info/

Given this, perhaps the most obvious justice and science protecting option would be some manner of licensing system for PEL occupation, with inbuilt time constraints or a periodic requirement for renewal (in order to block claims in perpetuity). This might be a stand-alone system with fairly limited ambitions (the securing of reasonably fair access among those technologically equipped to take advantage of the opportunity) or it might be tied into a more ambitious system of compensation (also covering parties who have a claim upon the PEL but lack the technological means to occupy them). Taxation would be one option, understood as a mechanism for compensation rather than a form of charity or aid. While the less ambitious option might provide adequate scientific protection the more ambitious option might be more consistent with an overall concern for distributive justice. One of the main objections to the more ambitious option would no doubt be that it risks compensating for harms that are not done. States that do not have lunar programs and are not in a position to develop them are prevented from occupying PELs by these facts alone and not by the lunar activities of other states and agencies. Yet, if pressed too forcefully, this consideration may be difficult to square with the affirmation by the OST that the Moon is the province of all mankind. Compensatory measures to promote access, and give more substantive content to the idea that the Moon is a commons, may then be appropriate.

Overall, compensation might be graduated to take account of how seriously disadvantaged a state was by PEL occupation by another state or by an agency responsible to that state. (Upon whom, in the absence of further alterations to space law, legal responsibility would ultimately still devolve.) Compensation might take the form of direct financial payment, debt relief or a transferable claim upon special access to some other scarce resource on the Moon or elsewhere. What matters here is not so much the actual form of the compensation but simply the fact that there are some obvious viable options for the latter. Each of them would have their own merits and disadvantages.

They would, for example, come with the familiar problems of all compensatory mechanisms, such as, firstly, establishing and sustaining a genuinely authoritative regulatory body which was not simply the mouthpiece for the larger terrestrial powers or else some nightmarish Weberian bureaucracy with its own priorities and set of dynamics. The problem of "who regulates the regulators" is familiar from other contexts. And secondly, there is the danger that compensatory mechanisms might work all too well, with the result that terrestrial agents who might otherwise favor lunar protection might find themselves effectively 'bought off'. A decision that there ought to be some form of compensatory mechanism would need to be the beginning of a much more detailed discussion.

Issues of justice arise not only in relation to nations that currently lack a space program (but who might develop one in the future when PEL occupancy might easily be 100%). They arise also in relation the possible loss of a PEL, should some manner of intrusive economic development of the Moon be approved. Who then should lose their claim upon a peak? Localized development might unfairly (even repeatedly) disadvantage one claimant or group of claimants while leaving others, elsewhere on the lunar surface, secure and unchallenged. (Analogously, there may be a case for the compulsory purchase of my house for the purposes of some socially useful development, but if the law is only ever exercised in relation to my properties and never anyone else's then I may reasonably claim that something has gone badly wrong.) Compensation mechanisms might be envisaged that allow for the creation of new PELs as a result of lunar development. This might create some concern about (3), i.e. lunar protection. But even if such could be satisfied, it is difficult to imagine that the possibility of access to a replacement site would avoid all disputes about fairness given that any new, replacement, PELs might not actually share all of the desirable and relational features of an old PEL (e.g. proximity to other resources) and might also not address special requirements for continuity of occupancy in one spot. In the light of this, the ideal solution would seem to involve both compensation and the giving of special weight to the views of the occupying party. However, giving too much weight to the latter might incentivize pre-emptive occupancy with a view towards later commercial development. A solution should then, perhaps, include the prevention of such development-blocking or development-enabling claims. Otherwise, agreement to development by an occupying country might be leveraged into an apparent sanction for such development. Accordingly, consideration (3), concerning planetary protection, might then be brought into play as a broader safeguard against strategic PEL occupation linked to the wider exploitation of lunar resources.

Overall, there may be no clear way to reduce the ethical complexity of these questions, but this may have more to do with the irreducible complexity of ethics than it does with the specifics or uniqueness of the circumstances under consideration. In everyday life and in ethical deliberation about the law, such complexity does not in practice leave us paralyzed or unable to establish decision procedures that yield an approximation of justice. On the plus side, the range of core ethical considerations which are in play does not look unmanageable or liable to place any excessively demanding constraints upon agents. The ethical considerations do not seem to be either moot or easily silenced within policy and legal discussions on purely pragmatic grounds. In fact, this situation invites a consideration of practical means to mitigate against the most egregious cases of strategic PEL occupation under the guise of scientific activity, to which we turn in the next section.

## 6. A CONCEPTUAL BASIS FOR PRACTICAL GUIDELINES FOR SCIENCE AT THE PELs

The PEL thought experiment reveals that the Moon's status as the "common province of all mankind" may clash with the practical realities of scientific experiments affecting high-value, scarce lunar resources, such as unique lunar features. As the experiment suggests, the rights of all parties to the OST to use, occupy, or control lunar features may be more accurately thought of as rights that are common in principle, but differentiated in practice.

The rights of 'early movers' who undertake scientific activities at valuable lunar features are differentiated for several reasons. States with greater spacefaring capabilities and that act earliest may seize the privilege and de facto right of using or occupying such lunar features to the practical exclusion of others. Beyond the effective right to exclude others, early movers may enjoy several other forms of advantage denied to their successors. Among these are the freedom and opportunity to use, occupy, and effectively control valuable lunar features in pursuit of scientific, exploratory, and perhaps even commercial interests with minimal interference.

Does this situation doom the Moon to become an ungoverned space where the mighty and quick prevail at the expense of the weak and slow?  In our reading of the treaty, it may be so.  In practice, however, a different outcome may yet prevail.

Parties to the OST are unlikely to be able to negotiate a new agreement to address the apparent loophole permitting strategic occupation of the PELs, but they may share an interest in developing other means to limit or stigmatize this conduct.  Rather than a new treaty, parties may seek other ways to create common understandings and standards of conduct that can guide responsible scientific activities at the Moon and create a basis for the articulation of grievances against those who conduct their scientific activities as a pretext to occupy valuable lunar features.  These common understandings need not be codified in an international agreement at first or ever.  Instead, they could evolve on the basis of relevant actors' expressed recognition of them and their practices.

Recognizing, as above, that at lunar features parties may in practice enjoy common but differentiated rights and responsibilities, we may extrapolate that early movers conducting scientific activities have responsibilities different from and greater than those of non-active parties because of the potential impact of their activities on other parties' interests.  What, then, constitutes the responsible conduct of scientific activities on the Moon?

Here, a large class of issues concerning the responsible conduct of science, and science in a democratic context, will be inherited [32, 33]. However, our principle concern here will be with specific problems for scientific practice posed by lunar activity. One way to approach this question is to infer from the values, interests, and principles expressed in the OST a set of practical criteria for evaluating the soundness and permissibility of scientific activities at the PELs.  Since a large class of scientific activities are likely to be unproblematic, these criteria would be useful mainly to evaluate those activities that appear to create an illicit basis for appropriation, occupancy, or effective control of a lunar feature.  These criteria would create a practical test for the permissibility of particular scientific experiments, rather than categorical proscriptions or prescriptions: scientific activities meeting all or most of the criteria would pass the test and be deemed responsible, while those failing to observe all or most of the criteria would fail the test and be deemed irresponsible.

This test would be useful not as a legal instrument or a set of hard and fast rules, but as a practical guide to good conduct – an outline of best practices or 'rules of the road' – for scientific activities at the PELs and elsewhere on the Moon.  Indeed, this approach would be consistent with a broader turn toward non-binding governance mechanisms taking place in various issue-areas.

Given this objective, let us posit that the responsible conduct of scientific activities on the Moon should observe two basic principles implied by and inferred from the OST: 'proportionality' and 'reasonable means.'

*Proportionality*
In international affairs, ranging from states' rights to self-defense to environmental governance, the ideal of 'proportionality' refers to a desirable balance of means and ends in state action. The means by which an act is carried out ought to be proportional to its ends. Action should be prudent and restrained, so that its overall impact does not exceed what is required to achieve its stated objective.

In the lunar context, any scientific activity is likely to entail modification or effective appropriation of lunar features to some degree. If the principle of proportionality is applied to such cases, then the impact of a scientific activity on a lunar feature should be proportional to its scientific objective.

As in many other areas, in the realm of lunar scientific activities, no overarching authority exists to adjudicate or evaluate this match of means to ends. Nevertheless, evaluations of scientific conduct are familiar to scientists. Given the collaborative nature of scientific practice and the partly self-governing character of many scientific communities, scientists' activities are routinely subject to evaluation by peer specialists in different forms. In fact, scientific activities may in fact be more conducive to evaluation against common international standards than other forms of international conduct. No obstacle prevents the application of such scientific judgments – however contingent, fluid, and imperfectly consensual they may be – to scientific activities conducted at valuable lunar features.

For our purposes, then, we may say that scientific activities on the Moon should be at least defensible to their relevant community of international scientific experts as having an impact that is commensurate with their scientific objective. If observed in practice, this criterion would prohibit at least the most impactful and large-scale activities from being undertaken under the pretext of scientific research.

*Reasonable means*
Applied to our setting, the principle of "reasonable means" would prescribe that parties conducting lunar scientific activities take reasonable means to minimize harms to and impacts upon other parties and maximize protections for other parties. Reasonable means are those which might fairly and properly be required of a party conducting scientific research on a lunar features. They should be suitable, just, and moderate.

In particular, parties should be able to demonstrate that, in the conduct of their scientific activities, they have taken reasonable means to 1) minimize the effective exclusion of other parties from the lunar feature or environment and 2) minimize the depletion, exhaustion, or other lasting impact of their activity on the lunar feature or its environment.

Moreover, parties should be able to demonstrate that, in the conduct of their scientific activity, they have taken reasonable means to 1) maximize the conservation of the lunar environment for other parties' scientific and exploratory purposes and 2) maximize precautions against environmental changes or degradations.

'Reasonable' is a relative and contextual term and its application is likely to be contestable. Yet, that the concept eludes a precise and universal definition does not prevent it from being used as a guide to judging conduct in diverse national and sub-national jurisdictions, including in definitions of criminal negligence. As a guide to practice, rather than a legal principle, the concept of "reasonable means" can refer to means justifiable to a knowledgeable and impartial third party, such as a community of international scientific experts with relevant expertise.

Even if we accept some plasticity in the application of the 'reasonable means' concept, scientific experts may still disagree over what it should mean in practice to an extent that would render it inapplicable. For example, is implementing a measure that will generate a 1% reduction in environmental impact reasonable if it costs 10 ten times more than proceeding without this measure? Can scientific activity in general proceed if expectations that such measures will be taken prevail? On such questions, reasonable agents can disagree and experts may be irreconcilably divided.

Still, if the objective is to devise a practical set of guidelines to reduce the likelihood of spurious scientific activities disguising lunar land grabs, then the concept of reasonable means may be sufficient to highlight at least the most gross or severe violations in a systematic way. This outcome alone would represent an improvement over the status quo.

In sum, if we regard early movers installing scientific experiments at the PELs to enjoy not only those rights to the Moon that they share on paper with other OST parties, but also differentiated privileges in practice, then we may determine that these early movers also bear differentiated responsibilities commensurate with their particular privileges. Practically and normatively, though not legally, these responsibilities might consist in duties to conduct their scientific activities in observance of the principles of proportionality and reasonable means. This means that scientific activities should create impacts upon other parties that are proportional to, and not in excess of, their stated scientific ends. It also means that actors should take reasonable means to minimize the harms and maximize protections to the lunar environment and to other parties in the course of their scientific activities. Elaborating guidelines for lunar scientific conduct based on these principles would create a common understanding of responsible behavior, establish a standard against which offenders' conduct could be evaluated, and provide a basis upon which other parties could exercise their OST-enshrined right to seek consultation with parties conducting such activities.

# 7. CONCLUSION

Given a degree of indeterminacy concerning what is and is not excluded by the OST (beyond the ruling out of explicit claims of "lunar property"), it seems that a case in defense of the legal appropriation of a pivotal lunar resource might be made. The "Peaks of Eternal Light", or other scarce areas, might readily be occupied for national prestige or for strategic commercial or political advantage and a legally-arguable precautionary claim upon non-interference by any rival power or agents could be underpinned by appeal to the value of scientific work undertaken. This being so, there is a case for a regulatory (and possibly a compensatory) framework. If conditions make it possible such a framework might be set out in an international agreement that might protect scientific enquiry and reduce the scope for any distributive injustice or resource monopolization which might result. However, at present, it seems unlikely that any such international agreement could be secured and so our best option may depend more upon other mechanisms to create common understandings and standards of conduct to guide scientific activities by appeal to notions of proportionality and reasonable means. As part of the process of shaping such mechanisms, deliberation about the peaks of eternal light may help to sharpen our understanding of questions about property rights in space and may help us to focus on creating workable policy solutions before some fait accompli shocks us into hasty action.

**Acknowledgements**

We thank Bob Bussey for permission to reproduce the figures. ME thanks the Aspen Center for Physics, funded by NSF grant # 1066293, for their hospitality while this paper was completed. AK gratefully acknowledges a Postdoctoral Fellowship from the Social Sciences and Humanities Research Council of Canada.